\documentclass[twocolumn,floatfix,showpacs,prd,aps,tightenlines]{revtex4}
\usepackage{graphicx}
\usepackage{psfrag}
\usepackage{amssymb}
\usepackage{dcolumn}
\usepackage{bm}

\newcommand{\bea}{\begin{eqnarray}}
\newcommand{\eea}{\end{eqnarray}}
\newcommand{\beq}{\begin{equation}}
\newcommand{\eeq}{\end{equation}}
\newcommand{\sYlm}[3]{{}_{#1}Y_{{#2}\,{#3}}}
\newcommand{\barsYlm}[3]{{}_{#1}{\bar Y}_{{#2}\,{#3}}}
\newcommand{\Real}{\ {\rm Re}}
\newcommand{\Imag}{\ {\rm Im}}

\newcommand{\scri}{\mathscr{I}}
\usepackage{mathrsfs}

\begin{document}

\title{A practical formula for the radiated angular momentum}

\author{Carlos O. Lousto}
\affiliation{Center for Computational Relativity and Gravitation,
School of Mathematical Sciences, Rochester Institute of Technology, 78
Lomb Memorial Drive, Rochester, New York 14623}

\author{Yosef Zlochower}
\affiliation{Center for Computational Relativity and Gravitation,
School of Mathematical Sciences, Rochester Institute of Technology, 78
Lomb Memorial Drive, Rochester, New York 14623}

\date{\today}

\begin{abstract}
We present a simple formula for the radiated angular momentum based
on a spin-weighted spherical harmonic decomposition of the Weyl scalar
$\psi_4$ representing outgoing radiation in the Kinnersley tetrad.
We test our formula by measuring the radiated angular momentum from
three simulations of non-spinning equal-mass black-hole binaries with orbital angular
momentum aligned along the $x$, $y$, and $z$ axes respectively. We
find that the radiated angular momentum agrees with the differences in the
remnant horizon spins and the initial angular momentum for each system.
\end{abstract}

 \maketitle

\section{Introduction}\label{Sec:Intro}

Recent breakthroughs in numerical relativity~\cite{Pretorius:2005gq,
Campanelli:2005dd,Baker:2005vv} have made it possible to accurately
evolve black-hole---black-hole~\cite{Campanelli:2006uy,Baker:2006kr,Buonanno:2006ui,Pfeiffer:2007yz} and
black-hole---neutron-star~\cite{Shibata:2006ks,Loffler:2006nu} binaries
for many orbits. The feasibility of evolving generic black-hole binaries was
demonstrated in~\cite{Campanelli:2007ew} where unexpectedly 
large recoil velocities of the merger remnant~\cite{Campanelli:2007ew,Campanelli:2007cg}
were found.
One of the important methods to
measure the accuracy of the full numerical evolutions is to monitor
the conservation of the energy, linear momentum,  and angular momentum. A
practical way of doing this in black-hole binary systems is to compute
the final black hole mass, linear momentum, and angular momentum from the
(isolated) horizon expressions, as well the radiated energy, linear momentum,
and angular momentum as measured far away from the system (formally at null infinity,
$\scri^+$). The sum of the corresponding remnant and radiated quantities
should converge to the initial mass, linear momentum, and angular momentum
of the system.

Expressions for the energy and momenta of an isolated system have long
been established (see
Refs.~\cite{Winicour_AMGR,Moreschi:2002ii} for reviews). Currently,
the
most popular method to extract gravitational radiation  is to
compute the Weyl scalar $\psi_4$ in a quasi-Kinnersley frame
~\cite{Campanelli:2005ia} or a frame that reduces to one
asymptotically. For practical purposes, it is then convenient to
express the radiation formulae for the mass, linear momentum, and angular
momentum in terms of $\psi_4$. That was done in
Ref.~\cite{Campanelli99}, see formulae (22)-(24) there, where the
radiated angular momentum was only given along the z-axis.  
We extend the computation of the radiated angular
momentum to all the three Cartesian components. These are useful to
verify the conservation of the angular momentum in generic black-hole
binaries, for instance, when spin precession is present
\cite{Campanelli:2006fy,Campanelli:2007ew,Campanelli:2007cg}.

\section{Derivation}\label{Sec:deriv}
The angular momentum on a space-like slice $\Sigma^+$ of $\scri^+$
(e.g.\ the intersection of a $u=t-r={\rm const}$ slice with
$\scri^+$) is given
by~\cite{Winicour_AMGR}
\begin{equation}
  \label{eq:link}
  J_{[i]} = \frac{1}{16 \pi} \Real \left\{ \oint_{\Sigma^+} 
   dS \xi^{A}_{[i]} \bar q_A \left[ 2 \psi^0_1 - 2 \sigma^0 \eth \bar
\sigma^0 - \eth(\sigma^0 \bar \sigma^0)\right] \right\},
\end{equation}
where the ${}^0$ superscript indicates the coefficient of the leading-order 
term in
an expansion in $1/r$, the spin-coefficient $\sigma$ and Weyl scalars $\psi_0,
\cdots, \psi_4$ are given in terms of a null tetrad 
$(l^a, n^a, m^a, \bar m^a)$
by (Eqs.~(3.7-3.8) of Ref.~\cite{Newman:1981fn}) 
\begin{eqnarray}
  \sigma &=& m^b m^a \nabla_a n_b, \\
  \psi_0 &=& -C_{a b c d} l^a m^b l^c m^d, \\
  \psi_1 &=& -C_{a b c d} l^a n^b l^c m^d, \\
  \psi_2 &=& -\frac{1}{2}\left(C_{a b c d} l^a n^b l^c n^d - C_{a b c d} l^a n^b m^c \bar m^d\right), \\
  \psi_3 &=& C_{a b c d} l^a n^b n^c \bar m^d,\\
  \psi_4 &=& -C_{a b c d} n^a \bar m^b n^c \bar m^d,
\end{eqnarray}
(where $C_{abcd}$ is the Weyl tensor),
and have the asymptotic form (Eq.~(4.4) in ~\cite{Newman:1981fn}):
\begin{eqnarray}
  \sigma &=& \sigma^0 r^{-2} + {\cal O}(r^{-3}),\\
  \psi_0 &=& \psi^0_0 r^{-5} + {\cal O}(r^{-6}),\\
  \psi_1 &=& \psi^0_1 r^{-4} + {\cal O}(r^{-5}),\\
  \psi_2 &=& \psi^0_2 r^{-3} + {\cal O}(r^{-4}),\\
  \psi_3 &=& \psi^0_3 r^{-2} + {\cal O}(r^{-3}),\\
  \psi_4 &=& \psi^0_4 r^{-1} + {\cal O}(r^{-2}),
\end{eqnarray}
 the coordinates on the
slice $\Sigma^+$ are such that the
volume element has the standard 
form for a 2-sphere, $dS=\sin\theta\, d\theta d\phi$,
the capital $A$ index ranges from $1$ to $2$ over the angular
coordinates in the slice, the dyad $q^A$ is the standard unit-sphere
dyad ($q^A = (1, i/\sin\theta)$ in $(\theta, \phi)$ coordinates),
the index $i$ ranges from $1$ to $3$ and represents the three Cartesian
components of the angular momentum, an overbar indicates complex conjugation,
the three $\xi^A_{[i]}$ are rotational Killing vectors intrinsic
to $\Sigma^+$, and the $\eth$ and $\bar \eth$ operators acting a
function $v_s$ of spin-weight $s$ have the form
\begin{eqnarray}
 \nonumber
 \eth v_s &=& -\partial_{\theta} v_s - \frac{i}{\sin\theta} \partial_\phi v_s + s v_s \cot\theta,\\
 \bar\eth v_s &=& -\partial_{\theta} v_s + \frac{i}{\sin\theta} \partial_\phi v_s - s v_s \cot\theta.
 \label{eq:eth_def}
\end{eqnarray}
The tetrad used in Eq.~(\ref{eq:link}) is partially fixed
by the choices $l_a = \nabla_a u$,
$u = (t - r)/\sqrt{2} + {\cal O}(\log r)$,
and $m^a = (0,0,1, i/\sin \theta)/\sqrt{2} + {\cal O}(1/r)$,
where $r$ is a luminosity distance (we consider here the
case where the spacetime and coordinates are such that the
metric approaches diag(1,1,1) at large $r$ and hence the coordinate
$r$ is a luminosity distance). Note
 that $l^a$ must
therefore be hypersurface orthogonal. The quasi-Kinnersley $l_a$, by contrast,
is not hypersurface orthogonal when the spin of the hole is non-vanishing and
will give an incorrect value for the angular momentum (e.g.\ zero for
a Kerr black hole). However,
as we will derive the radiated angular momentum in terms of $\psi_4$,
and the Kinnersley~\cite{Teukolsky73} $l_a$ differs from $\nabla_a u$ by
a type II rotation (which leaves $\psi_4$ invariant) and a trivial
boost of $\sqrt{2}$,
the Kinnersley $\psi_4$ can be used for the radiated angular momentum
formulae derived below since it is invariant under type II rotations.

Note that for any spin-zero function $f$ and Killing vector
(of the unit-sphere metric)
$\xi^A$,
\begin{equation}
\Real\left\{\oint_{\Sigma^+} dS \xi^A\bar q_A \eth f\right\}
 =\Real\left\{\oint_{\Sigma^+} dS \xi^A\bar q_A\,\frac{1}{2} \eth (f-\bar f)]\right\},
\label{eq:imaginary}
\end{equation}
i.e.\ only the imaginary part of $f$ contributes to the integral. Thus the final term
in Eq.~(\ref{eq:link}) does not contribute to $J$.

Let us define $\Phi_{[i]} = \xi^A_{[i]}q_A$ where $\xi^A_{[i]}$ are the standard
rotational Killing vectors about the $x$, $y$, and $z$ axes respectively,
then
\begin{eqnarray}
  \Phi_{[x]} &=& i \cos\theta \cos\phi - \sin \phi =
  2 i \sqrt{\frac{\pi}{3}}(\sYlm{-1}{1}{1} - \sYlm{-1}{1}{-1}),\nonumber\\
  \Phi_{[y]} &=& i \cos\theta \sin\phi + \cos \phi = 2 \sqrt{\frac{\pi}{3}} (\sYlm{-1}{1}{1} + \sYlm{-1}{1}{-1}),\nonumber\\
  \Phi_{[z]} &=& -i \sin\theta = -2 i \sqrt{\frac{2 \pi}{3}} \sYlm{-1}{1}{0},
\label{eq:Phi}
\end{eqnarray}
where the spin-weighted spherical harmonics have the form
\footnote{The definitions of $\eth$ and $\bar\eth$ in 
Eq.~(\ref{eq:eth_def}) differ from  the conventions in~\cite{Zlochower:2003yh}
by a factor of $-1$. We compensate for this factor in the definition
of $\sYlm{s}{\ell}{m}$ in Eq.~(\ref{eq:sYlm}). Thus the $\sYlm{s}{\ell}{m}$
used here are identical to those used in~\cite{Zlochower:2003yh}.}
\begin{eqnarray}
  \nonumber
  \sYlm{s}{\ell}{m} &=& (-1)^s\sqrt{\frac{(\ell-s)!}{(\ell+s)!}}
   \ \eth^s\ \sYlm{\!}{\ell}{m}\ \ \  \mbox{(for $s\geq0$)},\\
  \sYlm{s}{\ell}{m} &=& \sqrt{\frac{(\ell+s)!}{(\ell-s)!}}
   \ \bar\eth^{-s}\ \sYlm{\!}{\ell}{m}\ \ \  \mbox{(for $s\leq0$)}.
  \label{eq:sYlm}
\end{eqnarray}
Note that all three $\Phi_{[i]}$ can be expressed as a sum
of $\ell=1$ spin-weighted spherical harmonics (of spin-weight -1).
Hence the angular momentum loss is given by
\begin{equation}
  \label{eq:Jdot1}
\dot J_{[i]} = \frac{1}{16 \pi} \Real\left\{\oint_{\Sigma^+} dS \Phi_{[i]}
(2 \dot \psi^0_1 - 2 \dot \sigma^0 \eth {\bar\sigma}^0 - 2 \sigma^0\eth{\dot{\bar\sigma}}^0)\right\},
\end{equation}
where a dot indicates differentiation with respect to $u$.
Using the following relations~\cite{Newman:1981fn},
\begin{eqnarray}
  \dot \psi_1^0 &=&- \eth \psi_2^0 + 2 \sigma^0 \eth {\dot{\bar\sigma}}^0,\\
  \psi_2^0 - \bar\psi_2^0 &=& \bar\eth^2 \sigma^0 - \eth^2 \bar\sigma^0 + \bar \sigma^0 {\dot\sigma}^0 - \sigma^0 {\dot {\bar \sigma}}^0,\\
\psi_4^0 &=& - {\ddot {\bar \sigma}}^0,
\end{eqnarray}
Eq.~(\ref{eq:Jdot1}) reduces to (where we used result (\ref{eq:imaginary}))
\begin{eqnarray}
  \label{eq:Ldot_extra}
  \nonumber
\dot J_{[i]} &=& \frac{1}{16 \pi} \Real\left\{\oint_{\Sigma^+} dS \Phi_{[i]} 
 (\eth^3 \bar \sigma^0 - \eth\bar\eth^2\sigma^0- \right.\\
 &&\left. 3 \dot\sigma^0 \eth\bar \sigma^0 +
 3 \sigma^0 \eth\dot{\bar\sigma}^0 - \bar \sigma^0\eth \dot\sigma^0 +\dot {\bar\sigma}^0 \eth \sigma^0)\right\}.
\end{eqnarray}
Note that the first two terms do not contribute to $\dot J_{[i]}$
 since $\Phi_{[i]}$ contains only $\ell=1$ modes and $\sigma^0$ (which has
spin-weight 2) cannot contain $\ell=1$ modes. 
Hence we have
\begin{eqnarray}
  \label{eq:Ldot_NonKin}
  \nonumber
\dot J_{[i]} &=& \frac{1}{16 \pi} \Real\left\{\oint_{\Sigma^+} dS \Phi_{[i]} 
 (-3 \dot\sigma^0 \eth\bar \sigma^0 +
 3 \sigma^0 \eth\dot{\bar\sigma}^0 - \right.\\
 &&\left. \bar \sigma^0\eth \dot\sigma^0 +\dot {\bar\sigma}^0 \eth \sigma^0)\right\}.
\end{eqnarray}
In terms of the more standard choice of tetrad
with $l_a = \nabla_a (t-r)$, the radiated angular momentum takes the form
\begin{eqnarray}
  \label{eq:Ldot_Kin}
  \nonumber
\frac {d\,J_{[i]}}{dt} &=& \frac{1}{16 \pi}\Real\left\{\oint_{\Sigma^+} dS \Phi_{[i]}
 (-3 {N} \eth\bar {H} + 3 { H} \eth\bar { N} - \right. \\
 && \left. \bar { H}\eth { N} +\bar { N} \eth { H})\right\},
\end{eqnarray}
where $\partial_{tt} H = \partial_t N = \bar \psi_4$.

After expressing $H$ and $N$ in terms of spin-2 spherical harmonics
\begin{eqnarray}
  H &=& \sum_{\ell,m} H_{\ell\,m}\, \sYlm{2}{\ell}{m},\\
  N &=& \sum_{\ell,m} N_{\ell\,m}\, \sYlm{2}{\ell}{m},
\end{eqnarray}
Eq.~(\ref{eq:Ldot_Kin}) becomes
\begin{equation}
  \frac{d\,J_{[i]}}{dt} = -\frac{1}{16 \pi}\sum_{\ell,m,\ell',m'} \Real\{\bar H_{\ell\,m}N_{\ell'\,m'}
   d^{[i]}_{\ell\,m\,\ell'\,m'}\},
\end{equation}
where
\begin{equation}
d^{[i]}_{\ell\,m\,\ell'\,m'} = c^{[i]}_{\ell\,m\,\ell'\,m'}
 - \bar c^{[i]}_{\ell'\,m'\,\ell\,m},
\end{equation}
and
\begin{eqnarray}
\nonumber
c^{[i]}_{\ell\,m\,\ell'\,m'} &=& \oint_{\Sigma^+} dS \Phi_{[i]}(
   3 \sqrt{(\ell+2)(\ell-1)}
   \,\,\barsYlm{1}{\ell}{m}\,\, \sYlm{2}{\ell'}{m'}\\
 &&- \sqrt{(\ell'-2)(\ell'+3)}
  \,\,\barsYlm{2}{\ell}{m}\,\,\sYlm{3}{\ell'}{m'}).
\end{eqnarray}
The coefficients $d^{[i]}_{\ell\,m\,\ell'\,m'}$ 
have the simple form
\begin{eqnarray}
\nonumber
d^{[i]}_{\ell\,m\,\ell'\,m'} &=& \delta_{\ell\,\ell'}\Big[2 i \sqrt{\ell (\ell+1)-m m'}\, \delta_{|m-m'|\,1},\\
\nonumber
&&2  (m-m')\sqrt{\ell (\ell+1)-m m'}\delta_{|m-m'|\,1},\\
&&4 i m \delta_{m\,m'}\Big]
\label{eq:dlmlpmp}
\end{eqnarray}
\footnote{We have demonstrated that Eq.~(\ref{eq:dlmlpmp}) is correct
for $(\ell,m,\ell',m')$ in the range
$2\leq \ell \leq 10$, $-\ell \leq m \leq \ell$, $2 \leq \ell'\leq 10$
 and $-\ell'\leq m'\leq \ell'$, but have not proven it in general.}.
The radiated angular momenta then has the form
\begin{eqnarray}
  \nonumber
  \frac {d\,J_x}{dt} &=& \frac{1}{8 \pi} \sum_{\ell\,m}\left(
   \sqrt{\ell(\ell+1)-m(m+1)} \Imag\{\bar H_{\ell\,m} N_{\ell\, m+1}\}\right.\\
  &&\left.+\sqrt{\ell(\ell+1)-m(m-1)} \Imag\{\bar H_{\ell\,m} N_{\ell\, m-1}\}\right),
  \label{eq:Lxdot}\\
  \nonumber
  \frac{d\,J_y}{dt} &=& \frac{1}{8 \pi}\sum_{\ell\,m}\left(
   \sqrt{\ell(\ell+1)-m(m+1)} \Real\{\bar H_{\ell\,m} N_{\ell\, m+1}\}\right.\\
  &&\left.-\sqrt{\ell(\ell+1)-m(m-1)} \Real\{\bar H_{\ell\,m} N_{\ell\, m-1}\}\right),
  \label{eq:Lydot}\\
  \frac{d\, J_z}{dt} &=& \frac{1}{4 \pi} \sum_{\ell\,m}m \Imag\{\bar H_{\ell\, m} N_{\ell\, m}\}.
  \label{eq:Lzdot}
\end{eqnarray}
Note that if $\psi_4^0 = \sum P_{\ell\,m}\,\,\sYlm{-2}{\ell}{m}$,
then $N_{\ell\,m} = (-1)^m \int_{-\infty}^t \bar P_{\ell\,-m}(\tau) d\tau$,
and $H_{\ell\,m} =\int_{-\infty}^t N_{\ell\,m}(\tau)d\tau$.
In practice one needs to integrate from the starting time of the simulation
rather than $t=-\infty$. This presents no significant problem for
black-hole binary simulations as the error introduced by assuming no
radiation in the past is of the same order as the error introduced by using
initial data without the correct radiation content. In addition, 
these simulation typically start when the radiation is quite small compared
to the late-inspiral and merger waveforms.

\section{Numerical Tests}\label{Sec:test}
We tested the newly introduced formulae for the $x$, $y$, and $z$ components
of the radiated angular momentum by evolving the non-spinning,
quasi-circular S0 configuration
of Ref.~\cite{Campanelli:2006fg}  rotated such that orbital
angular momentum is aligned along the $x$ (S0X),  $y$ (S0Y), and
$z$ (S0Z) axes respectively.
For reference we provide the initial
data parameters in Table~\ref{table:ID} for these configurations.
The total radiated angular momentum for each configuration is reported
in Table~\ref{table:ID}. We ran the S0X and S0Y configurations with our
new AMR code~\cite{Campanelli:2007ew} with 9 levels of refinement
and central resolution of $M/40$. We extracted the waveform at radii $r=25M$,
$30M$, $35M$, and $40M$. We fit the calculated radiated angular
momentum to a linear and quadratic polynomial in $1/r$. We take the quadratic
extrapolation to be the value at $r=\infty$ and the quoted errors are
the differences between the linear and quadratic extrapolations.
 Note that we did not re-run the S0Z configuration,
but rather re-analyzed the waveform from an existing unigrid run with
a relatively low central resolution of $h=M/22.5$ and extraction
radii at $r=20M$, $25M$, and $30M$. In all cases we summed over
all modes with $\ell \leq 4$ \footnote{For these configuration
the $\ell=2$ modes dominate the contribution to the radiated
angular momentum.}.
The remnant hole for these configurations (SOX, S0Y, and S0Z all represent
the same binary) has spin $S_{\rm remnant}/M^2 = 0.639\pm0.001$ indicating
that the angular momentum loss is $\delta J/M^2 = 0.237\pm0.001$, which
agrees, to within the error estimates, with the radiated angular momenta
reported in Table~\ref{table:ID}.

\begin{table}
\caption{Initial data and total radiated angular momentum 
for quasi-circular, equal-mass black-hole binaries. The total
angular momentum is
$J/M^2 = 0.8764$ and the proper horizon separation is $l/M=10.01$.
 The punctures
are located at $\pm (X, Y, Z)$, with mass parameter $m_p/M = 0.4848$, 
and momentum $\pm(P_x, P_y, P_z)$.
 The orbital
parameter $(X, Y, Z)$ and $(P_x, P_y, P_z)$ are obtained by
the 3PN equations of motion for equal-mass binaries with orbital
frequency $M\Omega=0.0500$. The puncture
masses are obtained by setting the ADM mass of the system to
$(1.0000\pm0.0005)M$.
}
\begin{ruledtabular}
\begin{tabular}{llll}\label{table:ID}
Config &  S0X & S0Y & S0Z\\
$X/M$ & 0.000 & 0.000 & 3.280\\
$Y/M$ & 0.000 & 0.000 & 0.000\\
$Z/M$ & 3.280 & 3.280 & 0.000\\
$P_x/M$ & 0.0000 & 0.1336 & 0.0000\\
$P_y/M$ & -0.1336 & 0.0000 & 0.1336\\
$P_z/M$ & 0.0000 & 0.0000 & 0.0000\\
$J^x_{\rm rad}/M^2$ & $0.236\pm0.007$& $<10^{-7}$ & 0\\
$J^y_{\rm rad}/M^2$ & $< 10^{-6}$& $0.236\pm0.007$ & 0\\
$J^z_{\rm rad}/M^2$ & $< 10^{-6}$& $<10^{-7}$ & $0.24\pm0.01$\\
\hline
\end{tabular}
\end{ruledtabular}
\end{table}

\section{Conclusion}\label{Sec:conclude}

We used the Linkage definition of the angular momentum to
derive a formula for the radiated angular momentum
(\ref{eq:Ldot_Kin}), (\ref{eq:eth_def}), (\ref{eq:Phi}) in terms of
$\psi_4$. The remaining radiation formulae for the energy and linear
momentum have already been given in this form in
Ref.~\cite{Campanelli99}, Eqs. (22)-(23).

We then derived a simple formula for the radiated angular momentum based
on a spin-weighted spherical harmonic decomposition of $\psi_4$, and
have demonstrated that the formula provides the correct radiated angular
momentum for binaries that radiate along all three Cartesian directions.
We also compared those results with the spin of the horizon of the final black hole 
minus the total ADM angular momentum of the system, and find very good agreement.
Note that we also used these formulae to produce this kind of check in a generic
black-hole binary simulation involving unequal mass and unequal (precessing) spins,
the SP6 run in Ref.~\cite{Campanelli:2007ew}, and obtained satisfactorily agreement.
This paper thus provides practical formulae for direct application to extract
radiation information from current numerical relativity simulations of compact sources.

As a final point we note that there is an ambiguity in the definition of
angular momentum for generic asymptotically flat 
spacetimes~\cite{Winicour_AMGR,Moreschi:2002ii} that arises from the
supertranslation symmetry of $\scri^+$. This ambiguity can lead to inaccuracies
in the computed angular momentum for generic gauges. In general one needs to
confirm that the intrinsic metrics of the extraction spheres used to compute
 the modes of $\psi_4$ asymptotically approach
 $ds^2=r^2 d\theta^2+r^2\sin^2\theta d\phi^2$.

\acknowledgments
We thank Manuela Campanelli for providing her notes on
Ref.~\cite{Campanelli99} and Osvaldo Moreschi for enlightening discussions.
We gratefully acknowledge NSF for financial support
from grant PHY-0722315.
Computational resources were provided by the Funes cluster at UTB
and by the Lonestar cluster at TACC.

\bibliographystyle{apsrev}
\bibliography{../../Lazarus/bibtex/references}

\begin{thebibliography}{20}
\expandafter\ifx\csname natexlab\endcsname\relax\def\natexlab#1{#1}\fi
\expandafter\ifx\csname bibnamefont\endcsname\relax
  \def\bibnamefont#1{#1}\fi
\expandafter\ifx\csname bibfnamefont\endcsname\relax
  \def\bibfnamefont#1{#1}\fi
\expandafter\ifx\csname citenamefont\endcsname\relax
  \def\citenamefont#1{#1}\fi
\expandafter\ifx\csname url\endcsname\relax
  \def\url#1{\texttt{#1}}\fi
\expandafter\ifx\csname urlprefix\endcsname\relax\def\urlprefix{URL }\fi
\providecommand{\bibinfo}[2]{#2}
\providecommand{\eprint}[2][]{\url{#2}}

\bibitem[{\citenamefont{Pretorius}(2005)}]{Pretorius:2005gq}
\bibinfo{author}{\bibfnamefont{F.}~\bibnamefont{Pretorius}},
  \bibinfo{journal}{Phys. Rev. Lett.} \textbf{\bibinfo{volume}{95}},
  \bibinfo{pages}{121101} (\bibinfo{year}{2005}), \eprint{gr-qc/0507014}.

\bibitem[{\citenamefont{Campanelli
  et~al.}(2006{\natexlab{a}})\citenamefont{Campanelli, Lousto, Marronetti, and
  Zlochower}}]{Campanelli:2005dd}
\bibinfo{author}{\bibfnamefont{M.}~\bibnamefont{Campanelli}},
  \bibinfo{author}{\bibfnamefont{C.~O.} \bibnamefont{Lousto}},
  \bibinfo{author}{\bibfnamefont{P.}~\bibnamefont{Marronetti}},
  \bibnamefont{and}
  \bibinfo{author}{\bibfnamefont{Y.}~\bibnamefont{Zlochower}},
  \bibinfo{journal}{Phys. Rev. Lett.} \textbf{\bibinfo{volume}{96}},
  \bibinfo{pages}{111101} (\bibinfo{year}{2006}{\natexlab{a}}),
  \eprint{gr-qc/0511048}.

\bibitem[{\citenamefont{Baker et~al.}(2006{\natexlab{a}})\citenamefont{Baker,
  Centrella, Choi, Koppitz, and van Meter}}]{Baker:2005vv}
\bibinfo{author}{\bibfnamefont{J.~G.} \bibnamefont{Baker}},
  \bibinfo{author}{\bibfnamefont{J.}~\bibnamefont{Centrella}},
  \bibinfo{author}{\bibfnamefont{D.-I.} \bibnamefont{Choi}},
  \bibinfo{author}{\bibfnamefont{M.}~\bibnamefont{Koppitz}}, \bibnamefont{and}
  \bibinfo{author}{\bibfnamefont{J.}~\bibnamefont{van Meter}},
  \bibinfo{journal}{Phys. Rev. Lett.} \textbf{\bibinfo{volume}{96}},
  \bibinfo{pages}{111102} (\bibinfo{year}{2006}{\natexlab{a}}),
  \eprint{gr-qc/0511103}.

\bibitem[{\citenamefont{Campanelli
  et~al.}(2006{\natexlab{b}})\citenamefont{Campanelli, Lousto, and
  Zlochower}}]{Campanelli:2006uy}
\bibinfo{author}{\bibfnamefont{M.}~\bibnamefont{Campanelli}},
  \bibinfo{author}{\bibfnamefont{C.~O.} \bibnamefont{Lousto}},
  \bibnamefont{and}
  \bibinfo{author}{\bibfnamefont{Y.}~\bibnamefont{Zlochower}},
  \bibinfo{journal}{Phys. Rev. D} \textbf{\bibinfo{volume}{74}},
  \bibinfo{pages}{041501(R)} (\bibinfo{year}{2006}{\natexlab{b}}),
  \eprint{gr-qc/0604012}.

\bibitem[{\citenamefont{Baker et~al.}(2006{\natexlab{b}})}]{Baker:2006kr}
\bibinfo{author}{\bibfnamefont{J.~G.} \bibnamefont{Baker}} \bibnamefont{et~al.}
  (\bibinfo{year}{2006}{\natexlab{b}}), \eprint{gr-qc/0612117}.

\bibitem[{\citenamefont{Buonanno et~al.}(2006)\citenamefont{Buonanno, Cook, and
  Pretorius}}]{Buonanno:2006ui}
\bibinfo{author}{\bibfnamefont{A.}~\bibnamefont{Buonanno}},
  \bibinfo{author}{\bibfnamefont{G.~B.} \bibnamefont{Cook}}, \bibnamefont{and}
  \bibinfo{author}{\bibfnamefont{F.}~\bibnamefont{Pretorius}}
  (\bibinfo{year}{2006}), \eprint{gr-qc/0610122}.

\bibitem[{\citenamefont{Pfeiffer et~al.}(2007)}]{Pfeiffer:2007yz}
\bibinfo{author}{\bibfnamefont{H.~P.} \bibnamefont{Pfeiffer}}
  \bibnamefont{et~al.} (\bibinfo{year}{2007}), \eprint{gr-qc/0702106}.

\bibitem[{\citenamefont{Shibata and Uryu}(2006)}]{Shibata:2006ks}
\bibinfo{author}{\bibfnamefont{M.}~\bibnamefont{Shibata}} \bibnamefont{and}
  \bibinfo{author}{\bibfnamefont{K.}~\bibnamefont{Uryu}},
  \bibinfo{journal}{Phys. Rev.} \textbf{\bibinfo{volume}{D74}},
  \bibinfo{pages}{121503} (\bibinfo{year}{2006}), \eprint{gr-qc/0612142}.

\bibitem[{\citenamefont{Loffler et~al.}(2006)\citenamefont{Loffler, Rezzolla,
  and Ansorg}}]{Loffler:2006nu}
\bibinfo{author}{\bibfnamefont{F.}~\bibnamefont{Loffler}},
  \bibinfo{author}{\bibfnamefont{L.}~\bibnamefont{Rezzolla}}, \bibnamefont{and}
  \bibinfo{author}{\bibfnamefont{M.}~\bibnamefont{Ansorg}},
  \bibinfo{journal}{Phys. Rev.} \textbf{\bibinfo{volume}{D74}},
  \bibinfo{pages}{104018} (\bibinfo{year}{2006}).

\bibitem[{\citenamefont{Campanelli
  et~al.}(2007{\natexlab{a}})\citenamefont{Campanelli, Lousto, Zlochower, and
  Merritt}}]{Campanelli:2007ew}
\bibinfo{author}{\bibfnamefont{M.}~\bibnamefont{Campanelli}},
  \bibinfo{author}{\bibfnamefont{C.~O.} \bibnamefont{Lousto}},
  \bibinfo{author}{\bibfnamefont{Y.}~\bibnamefont{Zlochower}},
  \bibnamefont{and} \bibinfo{author}{\bibfnamefont{D.}~\bibnamefont{Merritt}}
  (\bibinfo{year}{2007}{\natexlab{a}}), \eprint{gr-qc/0701164}.

\bibitem[{\citenamefont{Campanelli
  et~al.}(2007{\natexlab{b}})\citenamefont{Campanelli, Lousto, Zlochower, and
  Merritt}}]{Campanelli:2007cg}
\bibinfo{author}{\bibfnamefont{M.}~\bibnamefont{Campanelli}},
  \bibinfo{author}{\bibfnamefont{C.~O.} \bibnamefont{Lousto}},
  \bibinfo{author}{\bibfnamefont{Y.}~\bibnamefont{Zlochower}},
  \bibnamefont{and} \bibinfo{author}{\bibfnamefont{D.}~\bibnamefont{Merritt}}
  (\bibinfo{year}{2007}{\natexlab{b}}), \eprint{gr-qc/0702133}.

\bibitem[{\citenamefont{Moreschi}(2004)}]{Moreschi:2002ii}
\bibinfo{author}{\bibfnamefont{O.~M.} \bibnamefont{Moreschi}},
  \bibinfo{journal}{Class. Quant. Grav.} \textbf{\bibinfo{volume}{21}},
  \bibinfo{pages}{5409} (\bibinfo{year}{2004}), \eprint{gr-qc/0209097}.

\bibitem[{\citenamefont{Winicour}(1980)}]{Winicour_AMGR}
\bibinfo{author}{\bibfnamefont{J.}~\bibnamefont{Winicour}}, in
  \emph{\bibinfo{booktitle}{General Relativity and Gravitation Vol 2}}, edited
  by \bibinfo{editor}{\bibfnamefont{A.}~\bibnamefont{Held}}
  (\bibinfo{publisher}{Plenum}, \bibinfo{address}{New York},
  \bibinfo{year}{1980}), pp. \bibinfo{pages}{71--96}.

\bibitem[{\citenamefont{Campanelli
  et~al.}(2006{\natexlab{c}})\citenamefont{Campanelli, Kelly, and
  Lousto}}]{Campanelli:2005ia}
\bibinfo{author}{\bibfnamefont{M.}~\bibnamefont{Campanelli}},
  \bibinfo{author}{\bibfnamefont{B.}~\bibnamefont{Kelly}}, \bibnamefont{and}
  \bibinfo{author}{\bibfnamefont{C.~O.} \bibnamefont{Lousto}},
  \bibinfo{journal}{Phys. Rev. D} \textbf{\bibinfo{volume}{73}},
  \bibinfo{pages}{064005} (\bibinfo{year}{2006}{\natexlab{c}}),
  \eprint{gr-qc/0510122}.

\bibitem[{\citenamefont{Campanelli and Lousto}(1999)}]{Campanelli99}
\bibinfo{author}{\bibfnamefont{M.}~\bibnamefont{Campanelli}} \bibnamefont{and}
  \bibinfo{author}{\bibfnamefont{C.~O.} \bibnamefont{Lousto}},
  \bibinfo{journal}{Phys. Rev. D} \textbf{\bibinfo{volume}{59}},
  \bibinfo{pages}{124022} (\bibinfo{year}{1999}), \eprint{gr-qc/9811019}.

\bibitem[{\citenamefont{Campanelli
  et~al.}(2006{\natexlab{d}})\citenamefont{Campanelli, Lousto, Zlochower,
  Krishnan, and Merritt}}]{Campanelli:2006fy}
\bibinfo{author}{\bibfnamefont{M.}~\bibnamefont{Campanelli}},
  \bibinfo{author}{\bibfnamefont{C.~O.} \bibnamefont{Lousto}},
  \bibinfo{author}{\bibfnamefont{Y.}~\bibnamefont{Zlochower}},
  \bibinfo{author}{\bibfnamefont{B.}~\bibnamefont{Krishnan}}, \bibnamefont{and}
  \bibinfo{author}{\bibfnamefont{D.}~\bibnamefont{Merritt}}
  (\bibinfo{year}{2006}{\natexlab{d}}), \eprint{gr-qc/0612076}.

\bibitem[{\citenamefont{Newman and Tod}(1980)}]{Newman:1981fn}
\bibinfo{author}{\bibfnamefont{E.~T.} \bibnamefont{Newman}} \bibnamefont{and}
  \bibinfo{author}{\bibfnamefont{K.~P.} \bibnamefont{Tod}}, in
  \emph{\bibinfo{booktitle}{General Relativity and Gravitation Vol 2}}, edited
  by \bibinfo{editor}{\bibfnamefont{A.}~\bibnamefont{Held}}
  (\bibinfo{publisher}{Plenum}, \bibinfo{address}{New York},
  \bibinfo{year}{1980}), pp. \bibinfo{pages}{1--36}.

\bibitem[{\citenamefont{{T}eukolsky}(1973)}]{Teukolsky73}
\bibinfo{author}{\bibfnamefont{S.~A.} \bibnamefont{{T}eukolsky}},
  \bibinfo{journal}{Astrophys. J.} \textbf{\bibinfo{volume}{185}},
  \bibinfo{pages}{635} (\bibinfo{year}{1973}).

\bibitem[{\citenamefont{Campanelli
  et~al.}(2006{\natexlab{e}})\citenamefont{Campanelli, Lousto, and
  Zlochower}}]{Campanelli:2006fg}
\bibinfo{author}{\bibfnamefont{M.}~\bibnamefont{Campanelli}},
  \bibinfo{author}{\bibfnamefont{C.~O.} \bibnamefont{Lousto}},
  \bibnamefont{and}
  \bibinfo{author}{\bibfnamefont{Y.}~\bibnamefont{Zlochower}},
  \bibinfo{journal}{Phys. Rev. D} \textbf{\bibinfo{volume}{74}},
  \bibinfo{pages}{084023} (\bibinfo{year}{2006}{\natexlab{e}}),
  \eprint{astro-ph/0608275}.

\bibitem[{\citenamefont{Zlochower et~al.}(2003)\citenamefont{Zlochower, Gomez,
  Husa, Lehner, and Winicour}}]{Zlochower:2003yh}
\bibinfo{author}{\bibfnamefont{Y.}~\bibnamefont{Zlochower}},
  \bibinfo{author}{\bibfnamefont{R.}~\bibnamefont{Gomez}},
  \bibinfo{author}{\bibfnamefont{S.}~\bibnamefont{Husa}},
  \bibinfo{author}{\bibfnamefont{L.}~\bibnamefont{Lehner}}, \bibnamefont{and}
  \bibinfo{author}{\bibfnamefont{J.}~\bibnamefont{Winicour}},
  \bibinfo{journal}{Phys. Rev.} \textbf{\bibinfo{volume}{D68}},
  \bibinfo{pages}{084014} (\bibinfo{year}{2003}), \eprint{gr-qc/0306098}.

\end{thebibliography}

\end{document}